\documentclass[epj]{svjour2}
\usepackage{graphics}
\usepackage{placeins}
\usepackage{hyperref}
\usepackage{marvosym} 
\usepackage{graphicx}
\usepackage{colordvi}
\usepackage{amssymb}
\usepackage{amsmath}
\usepackage{amsfonts}
\usepackage{epsfig}

\newcommand{\IGN}[1]{}
\newcommand{\ZWR}[1]{}

\begin{document}
%
\title{Drift reversal in asymmetric coevolutionary conflicts:\\Influence of microscopic processes and population size}
\titlerunning{Drift reversal in asymmetric coevolutionary conflicts}
\author{Jens Christian Claussen
}                     
%
%
\institute{Institut f{\"u}r Theoretische Physik und Astrophysik, Christian-Albrechts Universit{\"a}t, Olshausenstra{\ss}e 40, 24098 Kiel, Germany}
%
\date{%
May 10, 2007;
final version November 27, 2007
}
\abstract{
The coevolutionary dynamics in finite populations currently
is investigated in a wide range of disciplines, as
chemical catalysis, biological evolution, social and economic systems.
The dynamics of those systems can be 
formulated within the unifying framework of evolutionary game theory.
However it is not a priori clear which mathematical
description is appropriate when populations are 
not infinitely large.
Whereas the replicator equation approach describes
the infinite population size limit by deterministic 
differential equations, in finite populations the dynamics 
is inherently stochastic which can lead to new effects.
Recently, an explicit mean-field description in the form of a  
Fokker-Planck equation was derived for frequency-dependent selection
in finite populations based on microscopic processes.
In asymmetric conflicts between two populations with a cyclic dominance, 
a finite-size dependent drift reversal was demonstrated, depending on the 
underlying microscopic process of the evolutionary update.
Cyclic dynamics appears widely in biological coevolution,
be it within a homogeneous population, or
be it between disjunct populations as female and male.
Here explicit analytic address is given
and the average drift is calculated for
the frequency-dependent Moran process
and for different pairwise comparison processes.
It is explicitely shown that the drift reversal cannot 
occur if the process relies on payoff differences between
pairs of individuals.
Further, also a linear comparison with the average payoff
does not lead to a drift towards the internal fixed point.
Hence the nonlinear comparison function of the 
frequency-dependent Moran process, 
together with its usage of nonlocal information via the
average payoff, is the essential part of the mechanism.
\PACS{
      {87.23.-n}{Ecology and Evolution}   \and
      {89.65.-s}{Social and economic systems}
\vspace*{7.3mm}
     } 
} 
\maketitle

\section*{Introduction}
Biology offers a rich laboratory of
various types of oscillatory, chaotic and
stochastic dynamics.
Recently cyclic evolutionary dynamics 
has been observed 
in E.coli in vitro \cite{kerr}
and in vivo \cite{kirkup}, 
and attracted interest as a possible mechanism to
stabilize biodiversity.
This contributes to a long-standing debate how the emergence of 
new mutants is maintained in biological evolution:
Cyclic dynamics has been one of the
first proposals for such mechanisms \cite{hypercycle}.
Cyclic coevolution is not only observed in
asexual reproduction. 
A prominent observation 
of cyclic domination is observed
in side-blotched lizards \cite{lizards,Zam00}.
Three territorial mating behaviour strategies
of the male 
lizards occur, and coincide genetically
with orange, blue and yellow blotches.
While the cyclic dynamics 
of rock-paper-scissors type
can be demonstrated
without taking males and females into 
account explicitely,
hereby an improved quantitative understanding
is possible \cite{Sinervo01}.

Social and economic systems are a likewise 
interesting class of systems in which
cyclic dynamics
is observed. 
Social individua deciding 
in economic situations 
\cite{helbing92physicaa,helbing92physicaa193,helbing93physicaa196,helbing05pre,helbing94socialfield,helbing94pair,helbing96decision}
can fall into oscillatory cycles,
e.g.\ 
when loners, not participating in the game,
are added as a third strategy to a Prisoner's Dilemma
\cite{loners,szabohauert02}.
Evolutionary game theory is a unifying 
approach
for 
such systems
\cite{nowak06book,szaboreview,miekiszreview}.

In this paper, a paradigmatic cyclic game
is analyzed, which
is of likewise importance, for biological
mating behaviour, as well as in human 
social decision dynamics:
Dawkins' 
``Battle of the Sexes'' (BOTS),
a 2$\times$2 bimatrix game,
is
the simplest possible type of a cyclic game 
played by individuals between two homogeneous, mixed 
populations \cite{dawkins}.
In the mating behaviour of females and males, a
cyclical dominance of `slow' and `fast' strategies
can lead to a ``Battle of the Sexes'', or oscillations,
in the infinite population replicator dynamics
\cite{dawkins,hofbauer98,magurran}. 
The inherent stochasticity in a finite population
\cite{moran,nowak04,taylor04}
refines this picture, depending on population size
and underlying process
\cite{TCH05,CT05,TCH06,reichenbach06,CT06,C08}.

The aim of the paper is to
investigate in detail the drift reversal 
firstly reported in \cite{TCH05},
and to corrobate the simulations with
an analytic result,
allowing for a refined insight into the drift reversal.
The main results are 
(i) that for a generic class of pairwise comparison
processes a drift reversal does not occur, 
and (ii) that for the Moran
process,
the combination of
both 
a nonlinear reproductive fitness and the comparison with a
global function (average fitness)
are necessary for a drift reversal.

The paper is organized as follows. 
In Section \ref{sec:intro},
the BOTS payoff matrices
are introduced
and
the infinite population 
description of evolutionary game theory by the
replicator equation is recalled.
In Section \ref{sec:processes}, evolutionary birth-death 
processes are defined
in a unifying framework for comparison.
In Section \ref{sec:simu}
the influence of the stochasticity on the
time evolution of the population densities
is motivated by computer simulations of the process.

In the main Section \ref{sec:avgdrift},
the average drift 
(more formally introduced in Sect.\ \ref{subsec:defdriftreversal})
of the BOTS dynamics
is calculated explicitely
in finite populations, and
the population size corrections
are obtained
to first and second order 
analytically 
for four microscopic interaction
processes,
for neutral evolution as well
as for two linear processes, the
Local Update and the linearized Moran process.
The paper concludes by discussing and summarizing the results.

\section{
Battle of the Sexes: Replicator dynamics
\label{sec:intro}
}
This is the simplest cyclic game between two populations.
Following Dawkins \cite{dawkins}, the elementary payoffs read%
\IGN{
\begin{eqnarray}
\left(\begin{array}{c}
(\pi_A^\text{\Male},\pi_A^\text{\Female})
\\
(\pi_B^\text{\Male},\pi_B^\text{\Female})
\end{array}\right)
=
\left(\begin{array}{cc}
(+1,-1)  & (-1,+1) \\
(-1,+1) & (+1,-1)  
\end{array}\right)
\left(\begin{array}{c}
(j,i)\\(N-j,N-i)
\end{array}\right)
\end{eqnarray}
}
\begin{eqnarray}
\left(\begin{array}{c}
\pi_A^\text{\Female}
\\
\pi_B^\text{\Female}
\end{array}\right)
=
\left(\begin{array}{cc}
-1  & +1 \\
+1 & -1 
\end{array}\right)
\left(\begin{array}{c}
x\\1-x
\end{array}\right)
\end{eqnarray}
\begin{eqnarray}
\left(\begin{array}{c}
\pi_A^\text{\Male}
\\
\pi_B^\text{\Male}
\end{array}\right)
=
\left(\begin{array}{cc}
+1  & -1 \\
-1 & +1
\end{array}\right)
\left(\begin{array}{c}
y\\1-y
\end{array}\right)
\end{eqnarray}
for agents interacting with a population of
$(x,y,1-x,1-y)$
$=(i/N,j/N,(N-i)/N,(N-j)/N)$
agents in the strategies 
$(\pi_A^\text{\Male},\pi_A^\text{\Female},\pi_B^\text{\Male},\pi_B^\text{\Female})$.
As the opponent's 
payoff matrix is not the simple
transpose of the proponent's payoff matrix, 
such games are called
bimatrix games or asymmetric conflicts.
While in the replicator equations 
\cite{hofbauer98}
picture the
population cannot go extinct due do lack of discreteness, 
for the stochastic description in this paper the
population size will be fixed to $N$ female and $N$ male individuals.

For the relative frequencies, 
or abundance densities,
evolutionary game theory,
with the implicit assumption of an 
infinite population,
considers
the replicator equation
\begin{eqnarray}
\dot{x}&=&
x(\pi^A_\text{\Male}-\langle\pi_\text{\Male}\rangle)
\\
\dot{y}&=&
y(\pi^A_\text{\Female}-\langle\pi_\text{\Female}\rangle)
\end{eqnarray}
For the standard parameter choice 
(equivalent to ``Matching Pennies'')
of the BOTS
the replicator equation 
has a constant of motion
\begin{eqnarray}
H = - x (1-x) y (1-y).
\label{constantofmotion}
\end{eqnarray}
%
%
The replicator equations exhibits neutrally stable 
oscillations around the $x=y=1/2$ fixed point;
the adjusted replicator equation
(see \cite{hofbauer98})
has an attractive stable fixed point.
In so-called {\sl asymmetric conflicts} 
({\sl bimatrix games})
where members of the two (sub)populations 
can receive different payoffs, both populations
may gain different average payoffs, and the
denominators in the adjusted replicator equations
(being the proper $N\to\infty$ limit of the Moran 
process, see \cite{TCH06})
\begin{eqnarray}
\dot{x}&=&
\frac{
x(\pi^A_\text{\Male}-\langle\pi_\text{\Male}\rangle)
}{\frac{1-w}{w}+\langle\pi_\text{\Male}\rangle}
\\
\dot{y}&=&
\frac{
y(\pi^A_\text{\Female}-\langle\pi_\text{\Female}\rangle)
}{\frac{1-w}{w}+\langle\pi_\text{\Female}\rangle}
\end{eqnarray}
are different,
as
$\langle\pi_\text{\Male}\rangle=-\langle\pi_\text{\Female}\rangle$.
Thus the denominators cannot be absorbed
into a dynamical rescale of time (velocity transformation)
and both types of replicator equations not necessarily
exhibit the same stability properties.

How is this behaviour changed, and how far is it 
preserved in finite
populations?
This is the central question addressed in this paper.

\vspace{3mm}

\section{Evolutionary processes\label{sec:processes}}
To study dynamics in finite populations, 
it is advised to go down to the
microscopic interactions,
and to derive macroscopic equations
of motion herefrom,
eventually utilizing 
a finite-size expansion to
derive fluctuation corrections to the
deterministic limit.
As the microscopic dynamics may
depend on the system at hand,
the respective biological or behavioral
setup may require different interaction
and competition processes.
These, however, can be cast into
a unifying framework.
Following previous investigations
\cite{helbing92physicaa,TCH05},
 we consider
two classes of birth-death processes:
the frequency-dependent Moran process
\cite{moran,nowak04,taylor04} 
describing competition with the whole population,
and 
local two-particle interaction 
processes, 
with linear 
\cite{TCH05}
or Fermi-type 
\cite{szabohauert02,blume,TNP06}
dependence of the reproductive fitness as a function
of the payoff difference between two competing agents.
For all processes, the payoffs for an individual 
read
\\[1mm]
\begin{eqnarray}
\pi_A^\text{\Male} &=& 2y-1 
\mbox{~~~~~~}
\pi_A^\text{\Female} = 1-2x
\nonumber \\
\pi_B^\text{\Male} &=& 1- 2y 
\mbox{~~~~~~}
\pi_B^\text{\Female} = 2x-1.  
\label{payoffs}
\end{eqnarray}

\vspace{3mm}

\subsection{Moran process}
The Moran process \cite{moran}, a birth-death process, thus preserving $N$, 
is a standard model of mathematical genetics describing random
inheritance in overlapping generations.
In its original formulation, fitnesses of the genetic types
were independent of abundance densities in the population,
i.e., coevolution was not taken into account.
In the frequency-dependent Moran process \cite{nowak04,taylor04},
each individual competes with the whole population,
or an representative fraction of it, 
and reproduces proportional to this (cumulative) payoff
normalized by the average payoff in
the population.
With the payoffs averaged over the 
respective
population,
\begin{eqnarray}
\langle\pi^\text{\Male} \rangle  &=&  
\frac{i}{N} \pi_A^\text{\Male}
+\frac{N-i}{N} \pi_B^\text{\Male} 
\\
\langle\pi^\text{\Female}  \rangle &=& 
\frac{j}{N} \pi_A^\text{\Female}
+\frac{N-j}{N}\pi_B^\text{\Female} 
\end{eqnarray}
for given $(i,j)$ 
the transition probabilities of the 
possible four hopping events are given by%
\begin{eqnarray}
T^{+ \bullet}&=& 
\frac{1}{2}
\frac{1-w+w\pi_A^\text{\Male} }{1-w+w\langle\pi^\text{\Male}\rangle}
\frac{i}{N} \frac{N-i}{N}
\nonumber\\
T^{- \bullet}&=&
\frac{1}{2}
\frac{1-w+w\pi_B^\text{\Male} }{1-w+w\langle\pi^\text{\Male}\rangle}
\frac{i}{N} \frac{N-i}{N}
\nonumber\\
T^{\bullet +}&=&
\frac{1}{2}
\frac{1-w+w\pi_A^\text{\Female} }{1-w+w\langle\pi^\text{\Female}\rangle}
\frac{j}{N} \frac{N-j}{N}
\nonumber\\        
T^{\bullet -}&=& 
\frac{1}{2}
\frac{1-w+w\pi_B^\text{\Female} }{1-w+w\langle\pi^\text{\Female}\rangle}
\frac{j}{N} \frac{N-j}{N}
\nonumber
\end{eqnarray}
For  better comparison with the processes below,
an additional factor $1/2$ was introduced.
For this commonly used version of the 
frequency-dependent Moran process
one has to ensure that 
negative payoffs do not lead to
negative transition probabilities. 
This inconsistency is avoided 
by delimiting $w$ such that the denominator
remains positive.

\subsection{Local processes}
In many cases a competition of a single individual 
with the whole population may be unrealistic.
Therefore, it is advised to consider a local 
competition among individuals.
One process of this type is the Local update process
\cite{TCH05}, where one individual $b$ is selected randomly
for reproduction, 
compares with another randomly chosen individual $a$,
and changes strategy with probability
$\frac{1}{2}(1+w(\pi_a -\pi_b))$.

\begin{figure}[htbp]
\noindent
\epsfig{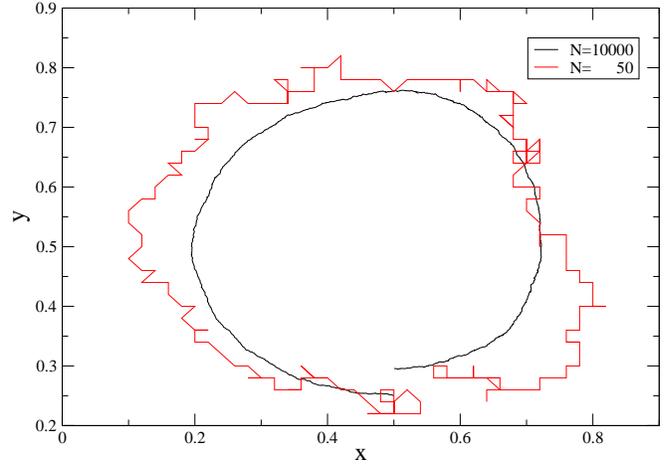} 
\caption{Stochastic motion around the fixed point,
Moran process ($w=0.5$) for different population
sizes.
Shown is a phase space plot, where $x$ denotes
the number of males in strategy A and $y$ denotes the
number of females in strategy A.
Due to the lack of helical paper, only one period of oscillation
is shown. 
One can observe that for the large population ($N=10000$) 
the trajectory spirals inwards towards the fixed point 
at $(\frac{1}{2},\frac{1}{2})$.
\label{fig_moran}}
\end{figure}

The frequency-dependent Moran process (MO), and the other processes below 
(LU = local update \cite{TCH05}, FP = Fermi process \cite{TNP06},
and a linearized Moran process (LM) considered below)
can more conveniently
be written by means of a reproductive function
\begin{eqnarray}
\label{phiMO}
\Phi_{\rm MO}(\pi_a,\langle\pi\rangle,w)&=& 
\frac{1}{2} \, \frac{1-w+w\pi_a }{1-w+w\langle\pi\rangle}\\
\label{phiLM}
\Phi_{\rm LM}(\pi_a,\langle\pi\rangle,w)&=&
\frac{1}{2}(1+w(\pi_a-\langle\pi\rangle))\\
\label{phiLU}
\Phi_{\rm LU}(\pi_a,\pi_b,w)&=&
\frac{1}{2}(1+w(\pi_a -\pi_b))\\
\Phi_{\rm FP}(\pi_a,\pi_b,w)&=& 
\frac{1}{1+\exp(-w (\pi_a-\pi_b))}
\label{Phi_fermi}
\label{phiFP}
\end{eqnarray}
so that the hopping probabilities are
\begin{eqnarray}
T^{ba}=\Phi (\pi_a,\pi_b,w) \frac{N_a}{N} \frac{N_b}{N}
\end{eqnarray}
for the pairwise comparison processes (LU,FP), and
\begin{eqnarray}
T^{ba}=\Phi (\pi_a,\langle\pi\rangle,w) \frac{N_a}{N} \frac{N_b}{N}
\end{eqnarray}
for the other processes (MO,LM) where an individual's payoff is 
compared to the payoff averaged over the whole population.

For all processes, $\Phi$ considers a two-particle (birth-death) process
where an individual with fitness $\pi_a$ compares 
(Eqs.\ \ref{phiMO}--\ref{phiLM}) with 
a representative sample of the population (i.e.\ the average fitness
$\langle\pi\rangle$), or
(Eqs. \ref{phiLU}--\ref{phiFP})
with another individual $\pi_b$.
The frequency-dependent Moran process
\cite{moran,nowak04},
the Local Update 
\cite{TCH05},
and the Fermi process 
\cite{TNP06}
are microscopic evolutionary processes 
discussed recently.
The Linearized Moran (LM) process arises 
as
approximation of the Moran (MO) process
in the weak selection limit $w\to 0$.

Note, that elsewhere for the Local update 
$w$ is replaced by $w/\Delta \pi_{\rm max}$,
and for the Moran process, $w$ may not exceed 
$1-\pi_{\rm min}$.
In the above notation, in the limit $w\to 0$ 
the Fermi process approaches the Local update 
process.

\begin{figure}[htbp]
\noindent
\epsfig{file=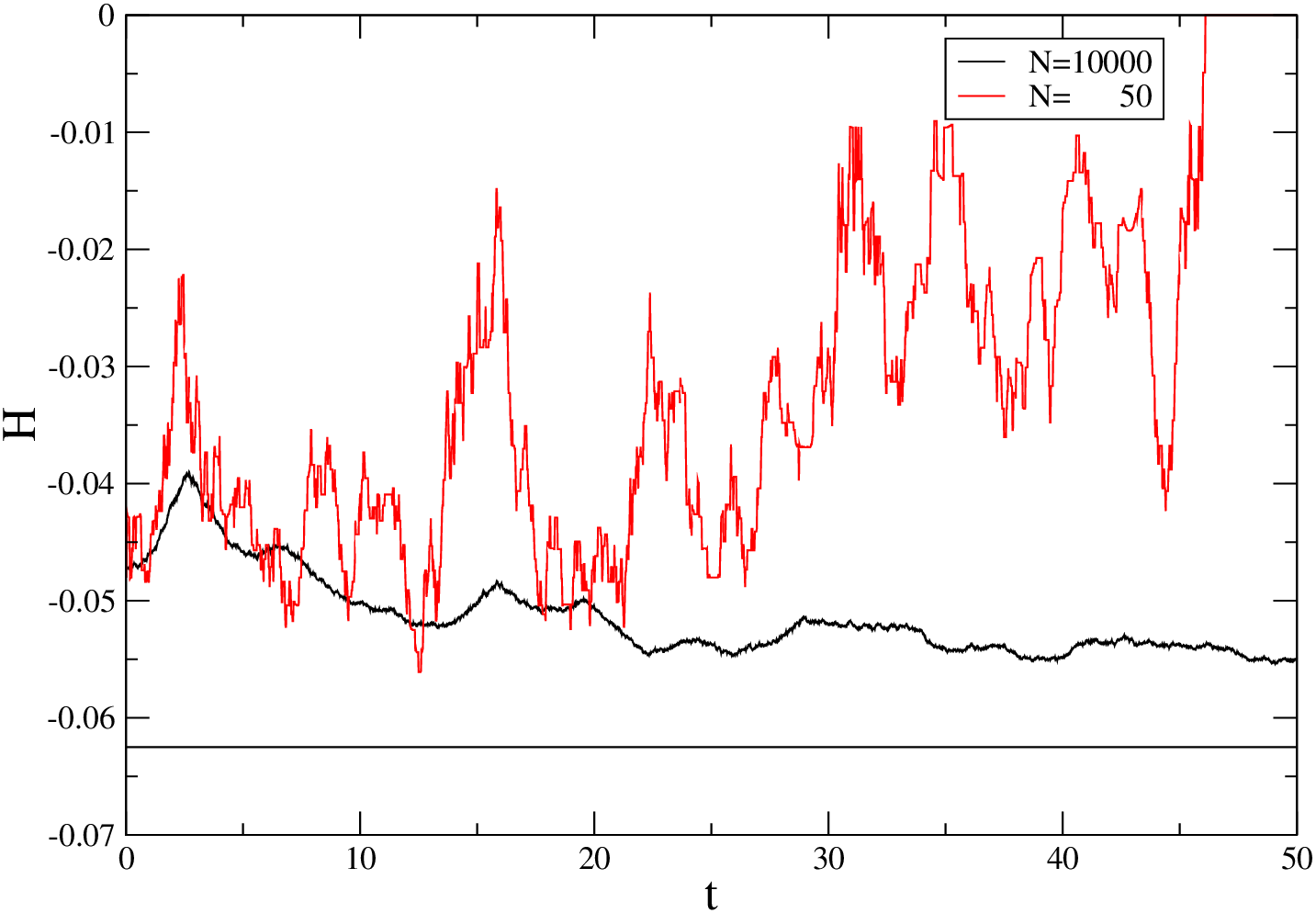,width=0.48\textwidth} 
\caption{Stochastic motion around the fixed point,
Moran process ($w=0.5$) for different population
sizes
(same as in Fig.\
\ref{fig_moran}).
Here for the initial trajectory
the value of $H$ 
(Eq.\ \ref{constantofmotion}) is shown as a function
of time. Time is measured in generation units,
i.e.\ $N$ update steps are performed per unit time so
that after $t=1$ each agent on average is updated once.
For $N=50$, $H$ fluctuates and also increases on average
(due to stochastic motion in the plane).
For $N=10000$, $H$ decays
as the trajectory approaches the fixed point.
For the corresponding deterministic system,
$H$ serves as a Lyapunov function of the
stable fixed point and admits the value $H_{\rm fix}=-1/16$
there (straight line).
\label{fig_moran2}}
\vspace{4mm}
\noindent
\centerline{\epsfig{file=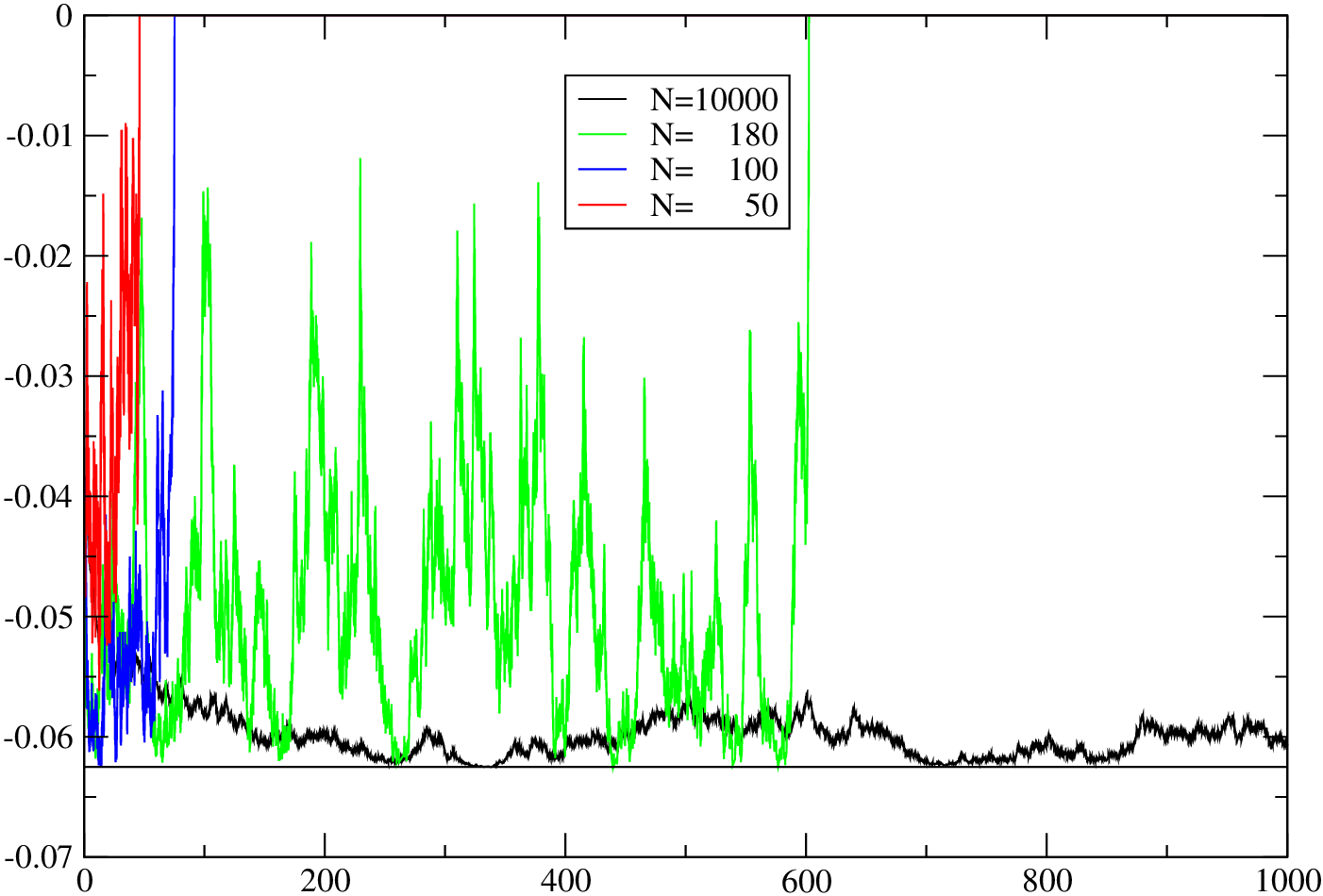,width=0.99\columnwidth}} 
\caption{Stochastic motion around the fixed point.
Moran process ($w=0.5$).
For small population sizes, 
the internal fluctuations are large compared to
the size of the phase space,
hence 
fixation to the border is reached
quickly. 
Above an intermediate range
(in which the process behaves similar to neutral stable fixed point),
fixation time diverges as 
the $N\to\infty$ behaviour of an
asymptotically stable fixed point is recovered.
\label{fig_moran4}}
\end{figure}

\section{Stochastic motion around the fixed point\label{sec:simu}}
Having defined different possible microscopic 
update processes, 
one desires to gain an intuitive understanding 
of the resulting stochastic motion, in
comparison of small (here, $N=50$) and large
(here, $N=10000$) populations.
Figure \ref{fig_moran}
compares the time evolution of the evolutionary trajectory
in phase space $(x,y)$ 
and
Figure \ref{fig_moran2}
in the time evolution of $H$.
For a large population, the qualitative dynamics of the
adjusted replicator equation is recovered,
i.e., convergence to the internal fixed point and
thereby decrease of $H$ which serves as Lyapunov function
in the deterministic limit.
The motion of the small population is comparatively more
stochastic, and after few generations, 
fixation to the border is reached.
Intuitively, the ``contracting force'' of the
stable fixed point is to weak
to ensure the observation
%
%
%
of a metastable state.
Figure
\ref{fig_moran4} 
shows a longer time interval and
four different population sizes.

\begin{figure}[htbp]
\noindent
\centerline{\epsfig{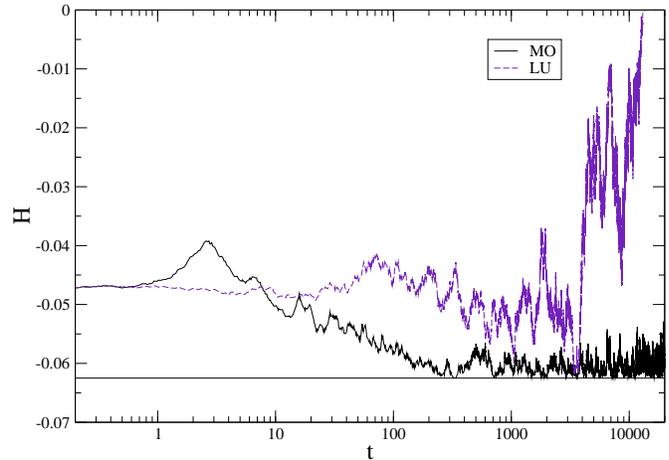}} 
\caption{Stochastic motion around the fixed point.
Time evolution over five decades for
the Moran process (black) 
and the Local update process (red/gray)
for a large population size $N=10000$
 ($w=0.5$).
For such large populations, the
stochastic motion is close to the
deterministic equations that follow
in the limit $N\to\infty$ \cite{TCH05}.
For the Local Update, the replicator equation
is neutrally stable at the fixed point,
and $H$ admits a constant of motion.
The stochastic realization shows zero drift in
$H$, but fluctuates, and ultimately
the process leads to fixation to the border.
Also the Moran process will reach
the border for $t\to\infty$,
however, after a long metastable transient.
For the Moran process, the
deterministic equation is the 
adjusted replicator equation which
has an asymptotocally stable fixed point.
Correspondingly $H$ decays 
describing a transient (metastable)
attractive behaviour towards the fixed point.
Note that due to the attracting force,
the motion keeps confined to 
the vicinity of the fixed point and
fluctuations are damped.
\label{fig_moranvslocal}}
\end{figure}


This paper addresses 
not only the influence of the population size, but also
the influence of
the microscopic update process.
Figure
\ref{fig_moranvslocal}
demonstrates that on longer time scales 
(here, the time scale is shown on a logarithmic scale
to show the convergence behaviour).
While the Local Update process performs a
specific type of random walk in $H$ (in the sense that the
internal fixed point is of neutral stability),
the Moran process 
first confines the motion in the vicinity of the
attracting internal point, 
but due to the discreteness of the process, 
fixates to the border in the limit $t\to\infty$.
The initial conditions are the same as in the
previous figures, starting from a state different from
the fixed point, to show that the Moran process
indeed leads to some type of convergence to the
metastable state.



\vspace{6mm}

\section{Deterministic and stochastic equations of motion 
at the level of the diffusion approximation\label{sec:avgdrift}}
In many cases population sizes are such small that 
deviations from a deterministic population density cannot
be neglected - especially because a species or genetic trait
can go extinct, or likewise, a new mutant can either extinct
or fixate (i.e., become the common trait in the whole population).
If populations are not too small, e.g.\ $N=10^2$, 
one however can think of a perturbation expansion in orders of
$1/N$ to derive stochastic corrections to the deterministic
equations of motion.
A classic approach in mathematical 
\clearpage\noindent
genetics in this direction is
the so-called diffusion approximation
\cite{crowkimura70,cherrywakely03},
which allows to derive fixation probabilities and densities from
a Kolmogorov forward, or Fokker-Planck, equation.
For a fixed population size and one trait
(higherdimensional cases are prohibitive),
the equivalence to a one-dimensional 
Markov chain
allows for a closed exact expression.

A similar framework applies for deriving the equations of
motion in zero and first order in $1/N$, and neglecting higher
orders; i.e., we operate at 
the level of the diffusion approximation.
Such approaches have been formulated in a wide
range of fields from genetics to behavioural dynamics
\cite{helbing92physicaa,helbing96decision,TCH05,TCH06,C08,crowkimura70,drosselreview}.
In the simplest setting of two strategies or traits,
from the hopping rates $T^{+},T^{-}$ we have 
\begin{eqnarray}
\dot{\rho}(x,t) = - \nabla (a(x) \rho(x,t)) + \nabla^{2}  (b^2(x)  \rho(x,t)) 
\end{eqnarray}
where $\nabla=\partial/\partial{}x$,
$a(x)=T^{+}-T^{-}$ and $b^2(x)=(T^{+}+T^{-})/N$
have to be derived for each microscopic update process
and payoff matrix.

\subsection{Stability and drift reversal
\label{subsec:defdriftreversal}}
The phase space average discussed in the next section 
can be interpreted as an approximation of the
time evolution of an observable $H$,
here, written for the case of two strategies only,
\begin{eqnarray}
\dot{H} \simeq
\langle \Delta H\rangle_{\rho x} 
&:=&
\!\!\int \!\! {\rm d} x \; \rho (x) \big[
(H(x+1/N) \!-\!H(x))T^{+} 
\nonumber
\\
&&
\hphantom{\int \!\! {\rm d} x \;  \rho (x) \big[} \!\!\!\!\!\!\!\!\!
+ (H(x-1/N) \!-\!H(x))T^{-}\big] ~~~~~
\end{eqnarray}
and, for an ensemble of arbitrary initial conditions,
we approximate $\rho(x)$ by a constant.
(Formally, one should distinguish between a continuous 
and a discrete phase space average; within this paper 
-- apart from constant prefactors -- the functions
for $H$ and the rates $T$ are continuous and differentiable.)

In the remainder we focus on functions $H$ which 
also are Lyapunov functions of an internal fixed point.
Then a decrease of $H$ can be interpreted as a motion
towards the fixed point, and an increase of $H$
is interpreted as an escape from the fixed point.
Thus we can define (omitting the subscript $x$ for the
phase space average)
\begin{eqnarray}
\langle \Delta H\rangle
&:=&
\!\!\int \!\! {\rm d} x  \big[
(H(x+
1/N
) \!-\!H(x))T^{+} 
\nonumber \\
&& \hspace*{1.4em}
+ (H(x-
1/N
) \!-\!H(x))T^{-}\big]
\end{eqnarray}
and define ``drift reversal'' as the change of sign
of 
$\langle \Delta H\rangle$.

If a change of system parameters ($N,w$, payoffs), 
leads to a changing sign of 
$\langle \Delta H\rangle$,
one observes the
fixed point gradually lose its stochastic
metastability. 
Due to this gradual transition, 
respective critical population
sizes are approximative and 
could be defined in (possibly many) different ways.
The essential advantage 
of assessing the stability from a sign reversal 
of this average drift $\langle \Delta H\rangle$ 
is that it can be calculated comparatively easy,
and in the case of the Battle of the Sexes, even analytically.

\section{Average drift: Battle of the Sexes}
\noindent
For the replicator equation,
Eq.\ (\ref{constantofmotion})
defines a constant of motion.
As we are interested in the finite-size corrections,
we can use $H$ as an observable for the distance 
to the interior fixed point.
For the processes defined above, the transition
probabilities allow to calculate 
the average change of the constant of motion
within the state space ($1 \leq i,j \leq N-1$)
as
\begin{eqnarray}
\nonumber
\langle \Delta H  \rangle 
&=&
\frac{1}{N^2(N-1)^{2}}
\sum_{i=1}^{N-1}
\sum_{j=1}^{N-1}
\Big[
\\&&
i(N-i)j(N-j)
(T^{+ \bullet}+T^{- \bullet}+T^{\bullet +}+T^{\bullet -})
\nonumber\\&&
- (i+1) (N-i-1)j (N-j) T^{+ \bullet}
\nonumber\\&&
- (i-1) (N-i+1)j (N-j) T^{- \bullet}
\nonumber\\&&
- i (N-i)(j+1) (N-j-1) T^{\bullet +}
\nonumber\\&&
- i (N-i)(j-1) (N-j+1) T^{\bullet -})
\Big]
\nonumber  
\\[6mm]
&=&
\nonumber
\frac{1}{N^2(N-1)^{2}}
\sum_{i=1}^{N-1}
\sum_{j=1}^{N-1}
\Big[
\\&&
j(N-j)
\big[
(2i-N) (T^{+ \bullet}-T^{- \bullet}) + (T^{+ \bullet}+T^{- \bullet})
\big]
\nonumber\\&&
\!\!\!\!\!
+i(N-i)\big[(2j-N)
(T^{\bullet +}-T^{\bullet -}) +(T^{\bullet +}+T^{\bullet -})
\big]\Big]
\nonumber  
\\[6mm]
&=&
\nonumber
\frac{N^2}{(N-1)^{2}}
\sum_{i=1}^{N-1}
\sum_{j=1}^{N-1}
\Big[
\\&&
y(1-y)
\big[
\frac{2x-1}{N}  (T^{+ \bullet}-T^{- \bullet}) \!+\!
\frac{1}{N^2}  (T^{+ \bullet}+T^{- \bullet})
\big]
\nonumber\\&&
\!\!\!\!\!
+x(1-x)\big[
\frac{2y-1}{N} (T^{\bullet +}-T^{\bullet -}) 
\!+\!\frac{1}{N^2} (T^{\bullet +}+T^{\bullet -})
\big]\Big].
\nonumber
\\
\end{eqnarray}
In the continuum limit
the sums
are replaced
 by integrals,
\begin{eqnarray}
\nonumber
\langle \Delta H  \rangle 
\!&=&\!
\int_{0}^{1} \! {\rm d}x 
\int_{0}^{1} \! {\rm d}y 
\Big[
\\&&
y(1-y)
\big[
\frac{2x-1}{N}  (T^{+ \bullet}-T^{- \bullet}) 
\!+\!
\frac{1}{N^2}  (T^{+ \bullet}+T^{- \bullet})
\big]
\nonumber
\\&&
\nonumber
\!\!\!\!\!
+x(1-x)\big[
\frac{2y-1}{N} (T^{\bullet +}-T^{\bullet -})
\!+\!\frac{1}{N^2} (T^{\bullet +}+T^{\bullet -})
\big]\Big]
\\[6mm]
&=&
\nonumber
\!\frac{1}{N}\!
\int_{0}^{1}\! {\rm d}x
\int_{0}^{1}\! {\rm d}y
x(1-x)y(1-y)
\big[
\\&&
\nonumber
\mbox{~~~~~~~~}
(2x-1)(\Phi_A^\text{\Male}-\Phi_B^\text{\Male})
+(2y-1)(\Phi_A^\text{\Female}-\Phi_B^\text{\Female})
\big]
\\&&
\nonumber
+\frac{1}{N^2}
\int_{0}^{1} {\rm d}x
\int_{0}^{1} {\rm d}y
x(1-x)y(1-y)\big[
\\&&
\mbox{~~~~~~~~~~~~~~~~~~~~~~~~~~~~~~~~}
(\Phi_A^\text{\Male}+\Phi_B^\text{\Male}+\Phi_A^\text{\Female}+\Phi_B^\text{\Female})
\big]
\nonumber
\\
\end{eqnarray}


%
%
%
\normalsize
 \clearpage\noindent
For given transition probabilities $T$ the drift of the
discrete process can be expressed exactly, and in the
case of linear processes, the resulting integrals 
are simple polynomial integrals.
\IGN{
\begin{eqnarray}
\langle \Delta H \rangle &=&
\frac{2}{N} \int_{0}^{1} {\rm d}x
\int_{0}^{1-x} {\rm d}y ( 
 y(x-z)(T^{RS}-T^{SR}) 
+x(z-y)(T^{SP}-T^{PS})
+z(y-x)(T^{PR}-T^{RP}) )
\nonumber
\\   
&&
-\frac{2}{N^2}\int_{0}^{1} {\rm d}x
\int_{0}^{1-x} {\rm d}y 
(y(T^{RS}+T^{SR})+x(T^{SP}+T^{PS})+z(T^{PR}+T^{RP})).
\end{eqnarray}
}
From this versatile expression, we can 
perform a comparison of the different processes.


%
\subsection{Neutral evolution and Local Update}
If $w=0$, or if all payoff elements are zero, 
all terms of type $(T^{+\bullet}-T^{-\bullet})$ vanish.
Now consider the case where the payoffs come into play.
For the Local Update, one has
\begin{eqnarray}
\Phi_A^\text{\Female} + \Phi_B^\text{\Female} =
\Phi_A^\text{\Male} + \Phi_B^\text{\Male} =
1 
\end{eqnarray}
and
\begin{eqnarray}
\Phi_A^\text{\Female} - \Phi_B^\text{\Female} &=& - 2w (2x-1) \\
\Phi_A^\text{\Male} - \Phi_B^\text{\Male} &=& 
+2w (2y-1).
\end{eqnarray}
\noindent
Thus the term
\begin{eqnarray}
\nonumber
(2x-1)(\Phi_A^\text{\Male}-\Phi_B^\text{\Male})
+(2y-1)(\Phi_A^\text{\Female}-\Phi_B^\text{\Female})
\end{eqnarray}
vanishes identically,
i.e.\ for the average drift
for the local update as well as for
neutral evolution only the $1/N^2$ order
survives,
\begin{eqnarray}
\nonumber
\langle \Delta H \rangle_{\rm LU}
&=&
\langle\Delta H \rangle_{\rm NE}
=
\frac{2}{N^2}
\int_{0}^{1} {\rm d}x \int_{0}^{1} {\rm d}y
x(1-x)y(1-y)
\nonumber
\\
&=&\frac{1}{18 N^2}. 
\label{driftneutral}
\end{eqnarray}
The discrete stochastic diffusion in this 
neutral case is equivalent to the genetic drift
of two independent types \cite{fisher,wright}
and the $1/N$ term vanishes, as expected.
But it is remarkable 
that the term in order $1/N$ does not
contribute for the Local Update process.
As this term corresponds for $N\to\infty$ to 
the ordinary replicator equation, 
this is fully consistent with $H$ being a
constant of motion.
For processes that depend on the payoff difference in
a nonlinear manner, as the Fermi process, 
$H$ is no longer a constant of motion.
However, the behaviour of  $\langle \Delta H \rangle$
is remarkable, and investigated for finite populations
in the next subsection.

\subsection{General nonlinear pairwise comparison processes}
Before approaching the nonlinear reproductive
fitness defined by the frequency-dependent Moran
process, 
it is instructive to analyze what happens if
one considers a general pairwise comparison
like for instance the Fermi process
(\ref{Phi_fermi}).
Let us assume the general case 
that the
difference between the reproductive 
functions 
$\Phi_A^\text{\Female} - \Phi_B^\text{\Female}$,
which is antisymmetric upon 
interchanging
the strategies $A,B$ of the individuals
anyhow,
merely is a 
function of the 
difference of the payoffs
\begin{eqnarray}
\Phi_A^\text{\Female} - \Phi_B^\text{\Female}
=:f(\pi_A^\text{\Female} - \pi_B^\text{\Female}),
\end{eqnarray}
which is fulfilled for the Local update,
the Fermi process, and practically all
common local comparison processes including
``Imitate if Better''.
Then $f$ is odd, $f(x)=-f(-x)$.
We also assume that this function is identical
for the female and male population -- 
this is justified as those differences should
be cast into the payoff matrix 
rather than manipulating the
reproductive functions.
For convenience, one can 
here require $f$ to be infinitely differentiable.
Then $f$ has 
only odd Taylor coefficients,
\begin{eqnarray}
\Phi_A^\text{\Female} - \Phi_B^\text{\Female}
&=& f(\pi_A^\text{\Female} - \pi_B^\text{\Female})
=f(\Delta\pi)
\nonumber
\\
\nonumber
&=& f^{'} \Delta\pi + \frac{f^{'''}}{3!} (\Delta\pi)^3
+\frac{f^{(5)}}{5!} (\Delta\pi)^5 + \ldots
\end{eqnarray}
The linear term is that of the Local update, 
whose cancellation in 
$\langle\Delta H\rangle$ we have seen before.
Upon inspection of, e.g., the  order $(2n+1)$ term,
one finds
apart from common factors
$H
{f^{(2n+1)}}/{(2n+1)!}$
that the integrand contains
\begin{eqnarray}
\nonumber
(2x-1)
(2(2y-1))^{2n+1} 
+
(2y-1)
(-2(2x-1))^{2n+1}
\\
=2^{2n+1} (2x-1)(2y-1)\big((2y-1)^{2n}-(2x-1)^{2n}\big)
\nonumber
\end{eqnarray}
That is, for every $n$,
the integral in order $1/N$ splits into the
difference between two integrals which are
symmetric in interchange of $x$ and $y$, 
thus identical, and the contributions of each
Taylor coefficient cancel.
Hence follows
\medskip
\medskip
\begin{theorem}
For a pairwise-type comparison process,
where the difference between 
reproductive fitnesses 
$\Phi_A-\Phi_B$
is an
infinitely differentiable function of the
payoff difference,
the 
average drift $\langle\Delta H\rangle$
in the
Battle of the Sexes (with the $\pm1$ payoff matrix)
is independent of the strength of
selection $w$ and equal to the drift 
of neutral evolution
(\ref{driftneutral}).
\end{theorem}
\medskip
\medskip
\medskip

As a corollary, there is no drift reversal
for the Local update 
as well as for the Fermi process,
confirming 
the numerical results in
\cite{TCH05,CT06}.

The result is remarkable in one sense.
Even though a nonlinear comparison process
can lead to a nonlinear replicator equation
-- which e.g.\ takes the form of a 
tangens hyperbolicus of the
payoff difference for the 
Fermi process \cite{TNP06} --
the average drift remains unaffected
even for strong selection and 
small populations.

\vspace*{4mm}

%
\subsection{Linearized Moran process \label{sec_linmoran}}
For the linearized Moran process, one has
\begin{eqnarray}
\Phi_A^\text{\Female} + \Phi_B^\text{\Female} =
1+\frac{w}{2}(\pi_A^\text{\Female} +\pi_B^\text{\Female}
-2 \langle \pi^\text{\Female}\rangle);
\end{eqnarray}
for the male population a corresponding equation holds.
The integrand in the diffusion (order $1/N^2$) term,
apart from a factor $H$, is
\begin{eqnarray}
2 + \frac{w}{2}(
 \pi_A^\text{\Female} +\pi_B^\text{\Female}
+\pi_A^\text{\Male} +\pi_B^\text{\Male}
-2 \langle \pi^\text{\Female}\rangle
-2 \langle \pi^\text{\Male}\rangle).
\nonumber
\end{eqnarray}
 \clearpage\noindent
As the sum of the payoffs $A$ and $B$
is zero for both populations,
and 
\begin{eqnarray}
\nonumber
\langle \pi^\text{\Female}\rangle= 
y(1-2x)+(1-y)(2x-1)=-(2x-1)(2y-1)
\\
=
-x(2y-1)+(1-x)(2y-1)=
-\langle \pi^\text{\Male}\rangle,
\nonumber
\end{eqnarray} 
again the game does not contribute 
to the diffusion term, 
which is, independently of $w$, identical to
the neutral case.
Now, does the game contribute to the first order?
Here,
\begin{eqnarray}
\Phi_A^\text{\Female} - \Phi_B^\text{\Female} &=& 
- \frac{w}{2} (\pi_A^\text{\Female}-\pi_B^\text{\Female})
= - w (2x-1) 
\\
\Phi_A^\text{\Male} - \Phi_B^\text{\Male} &=& 
+
 \frac{w}{2} (\pi_A^\text{\Male}-\pi_B^\text{\Male})
= 
+
w (2y-1)
\end{eqnarray}
and thus again the term
\begin{eqnarray}
\nonumber
(2x-1)(\Phi_A^\text{\Male}-\Phi_B^\text{\Male})
+(2y-1)(\Phi_A^\text{\Female}-\Phi_B^\text{\Female})
\end{eqnarray}
vanishes,
i.e.\ for the average drift
for the local update as well as for
neutral evolution only the $1/N^2$ order
survives,
\begin{eqnarray}
\nonumber
\langle \Delta H \rangle_{\rm LM}
&=&\frac{1}{18 N^2}. 
\label{driftlinearizedmoran}
\end{eqnarray}



%
\subsection{Moran process}
For the Moran process, 
it is possible to proceed 
even despite the 
nonlinearities.
The Moran payoffs are, e.g.\ for males in strategy A,
\begin{eqnarray}
\Phi^A_\text{\Male}
&=& \frac{1}{2}\frac{1-w+w\pi^A_\text{\Male}}{1-w+w\langle \pi_\text{\Male}\rangle}.
\nonumber
\end{eqnarray}
Using $\pi^A_\text{\Male}+\pi^B_\text{\Male}=0$,
and $\langle \pi_\text{\Female}\rangle+\langle \pi_\text{\Male}\rangle=0$,
it follows
\begin{eqnarray}
&&\hspace*{-2em}
\Phi^A_\text{\Male}
+
\Phi^B_\text{\Male}
+\Phi^A_\text{\Female}
+
\Phi^B_\text{\Female}
\\
\nonumber
&=&
\hphantom{+}
\frac{1}{2}\frac{(1-w+w\pi^A_\text{\Female})
(2-2w+w\pi^A_\text{\Male}+w\pi^B_\text{\Male})
}{(1-w+w\langle \pi_\text{\Male}\rangle)(1-w+w\langle \pi_\text{\Female}\rangle)}
\nonumber
\\&& 
+ \frac{1}{2}\frac{(1-w+w\pi^A_\text{\Male})
(2-2w+w\pi^A_\text{\Female}+w\pi^B_\text{\Female})
}{(1-w+w\langle \pi_\text{\Male}\rangle)(1-w+w\langle \pi_\text{\Female}\rangle)}
\nonumber
\\
&=&
\nonumber
\frac{2(1-w)^2}{(1-w)^2-w^2\langle \pi_\text{\Female}\rangle^2}.
\end{eqnarray}
In the weak selection limit, $w\to 0$,
this term approximates the constant 2.
For $w\to 1$, it vanishes asymptotically, but is 
negative for $\frac{1}{2}<w<1$, due to a pole at 
$w=\frac{1}{2}$.
While for $x=\frac{1}{2}$ or $y=\frac{1}{2}$ 
(a cross of lines in the state space) the average fitness
vanishes,
in the corners the square of the average fitness 
always is $1$.

\noindent
The difference terms in the reproductive functions are
\begin{eqnarray}
\Phi^A_\text{\Male}
-
\Phi^B_\text{\Male}
&=&
 \frac{1}{2}\frac{w(\pi^A_\text{\Male}-\pi^B_\text{\Male})}{1-w+w\langle \pi_\text{\Male}\rangle}
=
+ \frac{w(2y-1)}{1-w+w\langle \pi_\text{\Male}\rangle}
\nonumber
\\
\Phi^A_\text{\Female}
-
\Phi^B_\text{\Female}
&=&
\frac{1}{2}\frac{w(\pi^A_\text{\Female}-\pi^B_\text{\Female})}{1-w+w\langle \pi_\text{\Female}\rangle}
=
-\frac{w(2x-1)}{1-w+w\langle \pi_\text{\Female}\rangle}.
\nonumber
\end{eqnarray}
The $1/N$ term thus becomes,
observing $\langle \pi_\text{\Female}\rangle=-\langle \pi_\text{\Male}\rangle$,
\begin{eqnarray}
\nonumber
&&
\hspace*{-1em}
\frac{w}{N}
\int_{0}^{1} {\rm d}x \int_{0}^{1} {\rm d}y
x(1-x)y(1-y)
\\\nonumber&&\mbox{~~~~~~~~~~~~~~~}
\times \Big[
\frac{(2x-1)(2y-1)}{1-w+w\langle \pi_\text{\Male}\rangle}
-
\frac{(2x-1)(2y-1)}{1-w-w\langle \pi_\text{\Male}\rangle}
\Big]
\nonumber
\\[6mm]
\nonumber
&=&
\frac{w}{N}
\int_{0}^{1} {\rm d}x \int_{0}^{1} {\rm d}y
x(1-x)y(1-y) (2x-1)(2y-1)
\\\nonumber&&\mbox{~~~~~~~~~~~~~~~}
\times
\frac{
(1-w-w\langle \pi_\text{\Male}\rangle)-
(1-w+w\langle \pi_\text{\Male}\rangle)
}{(1-w)^2-w^2\langle \pi_\text{\Male}\rangle^2}
\\[6mm]
\nonumber
&=&
\frac{w}{N}
\int_{0}^{1} {\rm d}x \int_{0}^{1} {\rm d}y
x(1-x)y(1-y) (2x-1)(2y-1)
\\\nonumber&&\mbox{~~~~~~~~~~~~~~~}
\times
\frac{
-2w\langle \pi_\text{\Male}\rangle
}{(1-w)^2-w^2\langle \pi_\text{\Male}\rangle^2}.
\nonumber
\end{eqnarray}

\noindent
The average drift for the Moran process then reads
\begin{eqnarray}
\nonumber
\langle \Delta H \rangle_{\rm MO}
&=&
\int_{0}^{1} {\rm d}x \int_{0}^{1} {\rm d}y
\frac{x(1-x)y(1-y)}{(1-w)^2-w\langle\pi^\text{\Male}\rangle^2}
\\&&
\times \Big[
\frac{2(1-w)^2}{N^2}
-\frac{2w^2\langle\pi^\text{\Male}\rangle (2x-1)(2y-1)}{N}
\Big]
\nonumber
\\[6mm]
\nonumber
&=&
2 \int_{0}^{1} {\rm d}x \int_{0}^{1} {\rm d}y
\frac{x(1-x)y(1-y)}{(1-w)^2-w\langle\pi^\text{\Male}\rangle^2}
\\&&
\mbox{~~~~~~~~~~~~~~~~~}
\times 
\frac{2(1-w)^2 - Nw^2\langle\pi^\text{\Male}\rangle^2}{N^2}
\nonumber
\\[6mm]
\nonumber
&=&
\frac{2}{N^2} \int_{0}^{1} {\rm d}x \int_{0}^{1} {\rm d}y
x(1-x)y(1-y)
\\&&
\nonumber
\mbox{~~~~~~~~~~~~~~~~~~~~}
\times
\frac{(1-w)^2-Nw^2\langle\pi^\text{\Male}\rangle^2}{
(1-w)^2-w^2\langle\pi^\text{\Male}\rangle^2}.
\label{driftmoranexact}
\end{eqnarray}
For weak selection $w\ll \frac{1}{2}$, and $N \gg 1$,
one can approximate the last fraction, yielding
\begin{eqnarray}
\nonumber
\langle \Delta H \rangle_{\rm MO}
&\simeq &
\frac{2}{N^2} \int_{0}^{1} {\rm d}x \int_{0}^{1} {\rm d}y
x(1-x)y(1-y)
\\&&
\nonumber
\mbox{~~~~~~~~~~~~~~~~}
\times
[1-Nw^2(2x-1)^2(2y-1)^2)].
\end{eqnarray}

\vspace{2mm}

\subsection*{Weak selection limit of the drift for the
Moran process}
Despite reproductive success is based on a
comparison with the
average payoff (instead of a pairwise comparison), 
for the linearized Moran process
investigated
in Section \ref{sec_linmoran},
the drift reversal is lost.
Thus,
performing the weak selection limit 
at the level of the microscopic interaction
does not properly conserve the
properties of the average drift 
of the frequency-dependent Moran process.
But from the exact theory for the 
Moran case,
one can consider the approximation $w\to 0$
at a later stage:
\begin{eqnarray}
\nonumber
\langle \Delta H \rangle_{\rm MO}
&\simeq&
-\frac{1}{N}
\int_{0}^{1} {\rm d}x \int_{0}^{1} {\rm d}y
\,\,
2w^2 x(1-x)y(1-y)
\nonumber\\&&
\mbox{~~~~~~~~~~~~~~~~~~~~~~~~~}
\times (2x-1)^2 (2y-1)^2
\nonumber\\&&
+\frac{2}{N^2}
\int_{0}^{1} {\rm d}x \int_{0}^{1} {\rm d}y
x(1-x)y(1-y)
\nonumber\\
&=&
- \frac{2 w^2}{900 N} + \frac{1}{18 N^2}.
\label{driftmoranweakselection}
\end{eqnarray}
Both terms cancel
for  a $w$ that matches
\begin{eqnarray}
N_c=25/w^2.
\label{ncrit}
\end{eqnarray}
One should bear in mind that 
due to the
approximation $w \ll 1/2$ made above,
equations 
(\ref{driftmoranweakselection})
and 
(\ref{ncrit}) 
rely on the weak selection limit,
which is the biologically most relevant case.

An interesting viewpoint on its own 
is to consider
$w$ as a ``bifurcation'' parameter.
Conversely to a Hopf bifurcation,
where an oscillation amplitude 
grows 
 with the square root of
a bifurcation parameter,
here the effect of the Moran dynamics,
to stabilize the finiteness-stochastic
oscillations around the fixed point,
grows quadratically with
the strength of selection $w$.
Thus, for the 
fre\-quen\-cy-de\-pen\-dent Moran
process in finite populations,
the onset of the
fixed-point stabilizing mechanism
is a {\sl quadratic} effect 
of weak, but increasing, selection.

\vspace{9mm} 

\section{Discussion}
The investigation of the different update mechanisms
suggests the existence of two universality classes:
one, for which the internal fixed point is 
neutrally stable (in the $N\to\infty$ limit),
and a second, for which the internal fixed point is asymptotically
stable (in the $N\to\infty$ limit).

The Local update process as well as other 
(differentiable
and pair-symmetric) pairwise comparison processes,
as the Fermi process, and also a linearized Moran process,
belong to the first class.
The Moran process belongs to the second class,
and one may assume that 
more update processes 
(combining average fitness and a nonlinearity)
could be constructed for this class;
the necessary condition is that the
internal fixed point is attractive 
in the $N\to\infty$ limit.

\vspace{40mm}

In this sense, the properties that were known for the
deterministic replicator equations were recovered,
and for the Moran process class
the transition between the instable and (meta-) stable regime
can be described by calculating the 
drift reversal from $\langle \Delta H\rangle$.
This approach may be advantageous in high-dimensional
cases, where exact simulations of the process 
become costly; a phase space average of $\langle \Delta H\rangle$
is obtained much easier, and in very high dimensional spaces 
one may approximate by a Monte Carlo integration.

\vspace{13mm}

\section{Conclusions}
Cyclic coevolution 
in biological or social dynamics
is an interesting class of 
coevolutionary dynamics
and allows in an examplaric way to study 
stability, drift and diffusion
in finite populations.
Population sizes in markets,
decison processes, as well as in biological
populations from bacteria to ants and to vertebrates 
can vary by many orders of magnitude,
thus the influence of the
population size
by the stochasticity in finite populations 
is of fundamental relevance in all those
disciplines.

For
the simplest generic model of
a cyclic coevolution
in an asymmetric conflict game between two populations,
the ``Battle of the Sexes'',
analytical results for the average drift could be
obtained that support the
consistent understanding how the
different fixed point stability 
of ordinary and adjusted replicator
equations 
translate to finite populations. 

Partially the results are quite unexpected:
Not only for the Fermi process, but also for
a quite general class of pairwise comparison
processes
--~even when they lead to different 
replicator equations~-- 
the average drift is identical to the
neutral case, and equals
$1/(18N^2)$.
The second surprise is that,
for the average drift,
the linearized 
Moran process also falls into the
equivalence class of the neutral evolution.

But how does the drift reversal seen numerically
(and known in the $N\to\infty$ limit for the adjusted replicator equation)
then have to be explained?
Here one has to calculate the 
average drift for the Moran case,
using the 
{\sl nonlinear 
dependence on the average payoff}
to observe the drift reversal.
The hereby obtained average drift then
can be analyzed for weak selection,
resulting in an estimate of the
critical population size
which delimits the
stability regimes of ordinary and adjusted replicator
equations.

More generally speaking, 
one can conclude that,  
to stabilize the
coexistence of strategies in a 
large --~but finite~-- population, 
not only the global knowledge of the
overall success of the population
--~the average payoff~--
must be taken into account,
but in addition
it has to determine 
the reproductive fitness
in a nonlinear manner.
The (on process level) linearized Moran process,
as well as neutral evolution
or pairwise comparison,
leads to quicker extinction
of all but one of the four
strategy pairs.

\end{document}